\documentclass{svmult}
\usepackage{epsfig,graphicx} % figures
\usepackage[bottom]{footmisc}% places footnotes at page bottom
\usepackage{amssymb,amsbsy}  %RR for Douglas Gough
\usepackage{natbib,url}      %RR additions
\bibpunct{(}{)}{;}{a}{}{,}   %RR reference punctuation
\def\cite#1{\citealp{#1}}    %RR restore old astroncite \cite command
\def\authorindex#1{}  %RA to be redefined by editor at insertion into book
\def\figspath{.}  

\begin{document}\newcount\preprintheader\preprintheader=1

%RR file: rr-assp-defs.tex = extra ASSP definitions by Rob Rutten
%RR last: Mar 10 2009 
%RR note: %RR Rob-to-Rob    
%RR site: cp rr-assp-defs.tex ~/rr/tex/macros/.

\def\thisvolume{these proceedings}
%RR OOPS: no volume number and no page numbers, Springer muckup

%RR journal abbreviations
%%%%%%%%%%%%%%%%%%%%%%%%%
\def\aj{{AJ}}			
\def\araa{{ARA\&A}}		
\def\apj{{ApJ}}			
\def\apjl{{ApJ}}		
\def\apjs{{ApJS}}		
\def\ao{{Appl.\ Optics}} 
\def\apss{{Ap\&SS}}		
\def\aap{{A\&A}}		
\def\aapr{{A\&A~Rev.}}		
\def\aaps{{A\&AS}}		
\def\an{{Astron.\ Nachrichten}}
\def\aspcs{{ASP Conf.\ Ser.}}
\def\assp{{Astrophys.\ \& Space Sci.\ Procs., Springer, Heidelberg}}
\def\azh{{AZh}}			
\def\baas{{BAAS}}		
\def\jrasc{{JRASC}}	
\def\memras{{MmRAS}}		
\def\mnras{{MNRAS}}
\def\nat{{Nat}}		
\def\pra{{Phys.\ Rev.\ A}} 
\def\prb{{Phys.\ Rev.\ B}}		
\def\prc{{Phys.\ Rev.\ C}}		
\def\prd{{Phys.\ Rev.\ D}}		
\def\prl{{Phys.\ Rev.\ Lett.}} %RR	
\def\pasp{{PASP}}
\def\pasj{{PASJ}}		
\def\qjras{{QJRAS}}
\def\science{{Sci}}		
\def\skytel{{S\&T}}		
\def\solphys{{Solar\ Phys.}} 
\def\sovast{{Soviet\ Ast.}}  
\def\ssr{{Space\ Sci.\ Rev.}}
\def\svassp{{Astrophys.\ Space Sci.\ Procs., Springer, Heidelberg}}
\def\zap{{ZAp}}			
\let\astap=\aap
\let\apjlett=\apjl
\let\apjsupp=\apjs
\def\grl{{Geophys.\ Res.\ Lett.}}  %RR Weiss
\def\jgr{{J. Geophys.\ Res.}} %RR Manoharan

%RR astronomy and math commands copied from ASP
%%%%%%%%%%%%%%%%%%%%%%%%%%%%%%%%%%%%%%%%%%%%%%%
\def\ion#1#2{{\rm #1}\,{\uppercase{#2}}}  %RR ~>\, \sc > uc 
\def\deg{\hbox{$^\circ$}}
\def\sun{\hbox{$\odot$}}
\def\earth{\hbox{$\oplus$}}
\def\la{\mathrel{\hbox{\rlap{\hbox{\lower4pt\hbox{$\sim$}}}\hbox{$<$}}}}
\def\ga{\mathrel{\hbox{\rlap{\hbox{\lower4pt\hbox{$\sim$}}}\hbox{$>$}}}}
\def\sq{\hbox{\rlap{$\sqcap$}$\sqcup$}}
\def\arcmin{\hbox{$^\prime$}}
\def\arcsec{\hbox{$^{\prime\prime}$}}
\def\fd{\hbox{$.\!\!^{\rm d}$}}
\def\fh{\hbox{$.\!\!^{\rm h}$}}
\def\fm{\hbox{$.\!\!^{\rm m}$}}
\def\fs{\hbox{$.\!\!^{\rm s}$}}
\def\fdg{\hbox{$.\!\!^\circ$}}
\def\farcm{\hbox{$.\mkern-4mu^\prime$}}
\def\farcs{\hbox{$.\!\!^{\prime\prime}$}}
\def\fp{\hbox{$.\!\!^{\scriptscriptstyle\rm p}$}}
\def\micron{\hbox{$\mu$m}}
\def\onehalf{\hbox{$\,^1\!/_2$}}	
\def\onethird{\hbox{$\,^1\!/_3$}}
\def\twothirds{\hbox{$\,^2\!/_3$}}
\def\onequarter{\hbox{$\,^1\!/_4$}}
\def\threequarters{\hbox{$\,^3\!/_4$}}
\def\ubv{\hbox{$U\!BV$}}		
\def\ubvr{\hbox{$U\!BV\!R$}}		
\def\ubvri{\hbox{$U\!BV\!RI$}}		
\def\ubvrij{\hbox{$U\!BV\!RI\!J$}}		
\def\ubvrijh{\hbox{$U\!BV\!RI\!J\!H$}}		
\def\ubvrijhk{\hbox{$U\!BV\!RI\!J\!H\!K$}}		
\def\ub{\hbox{$U\!-\!B$}}		
\def\bv{\hbox{$B\!-\!V$}}		
\def\vr{\hbox{$V\!-\!R$}}		
\def\ur{\hbox{$U\!-\!R$}}

%%%%%%%%%%%%%%%%%%%%%%%%%%%%%%%%%%%%%%%%%%%%%%%%%%%%%%%%%%%%%%%%%%%%%%%%%%%%
%RR RJR additional commands
%%%%%%%%%%%%%%%%%%%%%%%%%%%%%%%%%%%%%%%%%%%%%%%%%%%%%%%%%%%%%%%%%%%%%%%%%%%%

%RR -- non-bullet item marker in itemize list 
\def\labelitemi{{\bf --}}  

%RR -- latin abbreviations
\def\rmit#1{{\it #1}}              %% italics (RR style, Kluwer)
\def\rmit#1{{\rm #1}}              %% redefine for ASP, A&A, ApJ, Springer
\def\etal{\rmit{et al.}}           %% use \etal\ for space behind it        
\def\etc{\rmit{etc.}}           
\def\ie{\rmit{i.e.,}}              %% , required (Webster 1681)
\def\eg{\rmit{e.g.,}}              %% , required (Webster 1681)
\def\cf{cf.}                       %% no Latin, always Roman (Webster 1686)
\def\viz{\rmit{viz.}}
\def\vs{\rmit{vs.}}

%RR -- mathematical
\def\rot{\hbox{\rm rot}}
\def\div{\hbox{\rm div}}
\def\lesssim{\mathrel{\hbox{\rlap{\hbox{\lower4pt\hbox{$\sim$}}}\hbox{$<$}}}}
\def\gtrsim{\mathrel{\hbox{\rlap{\hbox{\lower4pt\hbox{$\sim$}}}\hbox{$>$}}}}
\def\mathstacksym#1#2#3#4#5{\def#1{\mathrel{\hbox to 0pt{\lower 
    #5\hbox{#3}\hss} \raise #4\hbox{#2}}}}
\mathstacksym\lesssim{$<$}{$\sim$}{1.5pt}{3.5pt} % less than approximately
\mathstacksym\gtrsim{$>$}{$\sim$}{1.5pt}{3.5pt} % greater than approximately
\mathstacksym\lrarrow{$\leftarrow$}{$\rightarrow$}{2pt}{1pt} % equilibrium
\mathstacksym\lessgreat{$>$}{$<$}{3pt}{3pt} %% less or greater

\def\dif{\: {\rm d}}                       %% differential d with space
\def\ep{\:{\rm e}^}                        %% e^ with space and roman e
\def\dash{\hbox{$\,-\,$}}                  %% math-like hyphen
\def\is{\!=\!}                             %% = in text for tighter spacing

%RR --stellar stuff
\def\starname#1#2{${#1}$\,{\rm {#2}}}  %% \starname{\alpha}{Cen~A} 
\def\Teff{\hbox{$T_{\rm eff}$}}   

%RR -- units (in addition to the ASP ones above)
\def\kms{\hbox{km$\;$s$^{-1}$}}
\def\ms{\hbox{m$\;$s$^{-1}$}}
\def\Mxcm{\hbox{Mx\,cm$^{-2}$}}    %% no 2, damn tex

%RR -- magnetic field 
\def\Bapp{\hbox{$B_{\rm app}$}}    %% apparent flux density, Lites convention

%RR -- oscillations
\def\komega{($k, \omega$)}                 %% k - omega 
\def\kf{($k_h,f$)}                         %% f - k_h
\def\VminI{\hbox{$V\!\!-\!\!I$}}           %% V-I
\def\IminI{\hbox{$I\!\!-\!\!I$}}           %% I-I
\def\VminV{\hbox{$V\!\!-\!\!V$}}           %% V-V
\def\Xt{\hbox{$X\!\!-\!t$}}                %% X-t

%RR -- atomic levels
%%      use:    \level 3s3p 3Pe
%%              \level 3s$^2$ {1,3}P{e,o}
%%              \level {} 3Ge
\def\level #1 #2#3#4{$#1 \: ^{#2} \mbox{#3} ^{#4}$}   

%RR -- some spectral species
\def\specchar#1{\uppercase{#1}}    %% to be redefined for A&A = \sc
\def\AlI{\mbox{Al\,\specchar{i}}}  %% use \AlI\ for space behind it
\def\BI{\mbox{B\,\specchar{i}}} 
\def\BII{\mbox{B\,\specchar{ii}}}  
\def\BaI{\mbox{Ba\,\specchar{i}}}  
\def\BaII{\mbox{Ba\,\specchar{ii}}} 
\def\CI{\mbox{C\,\specchar{i}}} 
\def\CII{\mbox{C\,\specchar{ii}}} 
\def\CIII{\mbox{C\,\specchar{iii}}} 
\def\CIV{\mbox{C\,\specchar{iv}}} 
\def\CaI{\mbox{Ca\,\specchar{i}}} 
\def\CaII{\mbox{Ca\,\specchar{ii}}} 
\def\CaIII{\mbox{Ca\,\specchar{iii}}} 
\def\CoI{\mbox{Co\,\specchar{i}}} 
\def\CrI{\mbox{Cr\,\specchar{i}}} 
\def\CriI{\mbox{Cr\,\specchar{ii}}} 
\def\CsI{\mbox{Cs\,\specchar{i}}} 
\def\CsII{\mbox{Cs\,\specchar{ii}}} 
\def\CuI{\mbox{Cu\,\specchar{i}}} 
\def\FeI{\mbox{Fe\,\specchar{i}}} 
\def\FeII{\mbox{Fe\,\specchar{ii}}} 
\def\FeIX{\mbox{Fe\,\specchar{ix}}}
\def\FeX{\mbox{Fe\,\specchar{x}}}
\def\FeXVI{\mbox{Fe\,\specchar{xvi}}}
\def\FrI{\mbox{Fr\,\specchar{i}}}
\def\HI{\mbox{H\,\specchar{i}}} 
\def\HII{\mbox{H\,\specchar{ii}}} 
\def\Hmin{\hbox{\rmH$^{^{_{\scriptstyle -}}}$}}      %% H^min, elegant
\def\Hemin{\hbox{{\rm He}$^{^{_{\scriptstyle -}}}$}} %% He^min, idem
\def\HeI{\mbox{He\,\specchar{i}}} 
\def\HeII{\mbox{He\,\specchar{ii}}} 
\def\HeIII{\mbox{He\,\specchar{iii}}} 
\def\KI{\mbox{K\,\specchar{i}}} 
\def\KII{\mbox{K\,\specchar{ii}}} 
\def\KIII{\mbox{K\,\specchar{iii}}} 
\def\LiI{\mbox{Li\,\specchar{i}}} 
\def\LiII{\mbox{Li\,\specchar{ii}}} 
\def\LiIII{\mbox{Li\,\specchar{iii}}} 
\def\MgI{\mbox{Mg\,\specchar{i}}} 
\def\MgII{\mbox{Mg\,\specchar{ii}}} 
\def\MgIII{\mbox{Mg\,\specchar{iii}}} 
\def\MnI{\mbox{Mn\,\specchar{i}}} 
\def\NI{\mbox{N\,\specchar{i}}}
\def\NIV{\mbox{N\,\specchar{iv}}}
\def\NaI{\mbox{Na\,\specchar{i}}}
\def\NaII{\mbox{Na\,\specchar{ii}}}
\def\NaIII{\mbox{Na\,\specchar{iii}}}
\def\NeVIII{\mbox{Ne\,\specchar{viii}}} 
\def\NiI{\mbox{Ni\,\specchar{i}}} 
\def\NiII{\mbox{Ni\,\specchar{ii}}}
\def\NiIII{\mbox{Ni\,\specchar{iii}}} 
\def\OI{\mbox{O\,\specchar{i}}} 
\def\OVI{\mbox{O\,\specchar{vi}}}
\def\RbI{\mbox{Rb\,\specchar{i}}} 
\def\SII{\mbox{S\,\specchar{ii}}} 
\def\SiI{\mbox{Si\,\specchar{i}}} 
\def\SiII{\mbox{Si\,\specchar{ii}}} 
\def\SrI{\mbox{Sr\,\specchar{i}}}
\def\SrII{\mbox{Sr\,\specchar{ii}}}
\def\TiI{\mbox{Ti\,\specchar{i}}} 
\def\TiII{\mbox{Ti\,\specchar{ii}}} 
\def\TiIII{\mbox{Ti\,\specchar{iii}}} 
\def\TiIV{\mbox{Ti\,\specchar{iv}}} 
\def\VI{\mbox{V\,\specchar{i}}} 
\def\HtwoO{\mbox{H$_2$O}}        %% H2O %RR TeX doesn't accept numbers alas
\def\Otwo{\mbox{O$_2$}}          %% O2

%RR -- hydrogen spectrum features
\def\Halpha{\mbox{H\hspace{0.1ex}$\alpha$}} %% \Halpha\ for space behind it
\def\Ha{\mbox{H\hspace{0.2ex}$\alpha$}}
\def\Hbeta{\mbox{H\hspace{0.2ex}$\beta$}}
\def\Hgamma{\mbox{H\hspace{0.2ex}$\gamma$}}
\def\Hdelta{\mbox{H\hspace{0.2ex}$\delta$}}
\def\Hepsilon{\mbox{H\hspace{0.2ex}$\epsilon$}}
\def\Hzeta{\mbox{H\hspace{0.2ex}$\zeta$}}
\def\Lyalpha{\mbox{Ly$\hspace{0.2ex}\alpha$}}
\def\Lybeta{\mbox{Ly$\hspace{0.2ex}\beta$}}
\def\Lygamma{\mbox{Ly$\hspace{0.2ex}\gamma$}}
\def\Lycont{\mbox{Ly\hspace{0.2ex}{\small cont}}}
\def\Baalpha{\mbox{Ba$\hspace{0.2ex}\alpha$}}
\def\Babeta{\mbox{Ba$\hspace{0.2ex}\beta$}}
\def\Bacont{\mbox{Ba\hspace{0.2ex}{\small cont}}}
\def\Paalpha{\mbox{Pa$\hspace{0.2ex}\alpha$}}
\def\Bralpha{\mbox{Br$\hspace{0.2ex}\alpha$}}

%RR -- Na D
\def\NaD{\mbox{Na\,\specchar{i}\,D}}    %% use \NaD\ for space behind it
\def\NaDone{\mbox{Na\,\specchar{i}\,\,D$_1$}}
\def\NaDtwo{\mbox{Na\,\specchar{i}\,\,D$_2$}}
\def\NaID{\mbox{Na\,\specchar{i}\,\,D}}
\def\NaIDone{\mbox{Na\,\specchar{i}\,\,D$_1$}}
\def\NaIDtwo{\mbox{Na\,\specchar{i}\,\,D$_2$}}
\def\Done{\mbox{D$_1$}}
\def\Dtwo{\mbox{D$_2$}}

%RR -- Mg b 
\def\Mgbone{\mbox{Mg\,\specchar{i}\,b$_1$}}
\def\Mgbtwo{\mbox{Mg\,\specchar{i}\,b$_2$}}
\def\Mgbthree{\mbox{Mg\,\specchar{i}\,b$_3$}}
\def\MgIb{\mbox{Mg\,\specchar{i}\,b}}
\def\MgIbone{\mbox{Mg\,\specchar{i}\,b$_1$}}
\def\MgIbtwo{\mbox{Mg\,\specchar{i}\,b$_2$}}
\def\MgIbthree{\mbox{Mg\,\specchar{i}\,b$_3$}}

%RR -- Ca II H & K 
\def\CaIIK{\mbox{Ca\,\specchar{ii}\,K}}       %% use \CaIIK\ for space
\def\CaIIH{\mbox{Ca\,\specchar{ii}\,H}}
\def\CaIIHK{\mbox{Ca\,\specchar{ii}\,H\,\&\,K}}
\def\HK{\mbox{H\,\&\,K}}
\def\Kthree{\mbox{K$_3$}}      %% numbers not permitted, alas
\def\Hthree{\mbox{H$_3$}}
\def\Ktwo{\mbox{K$_2$}}
\def\Htwo{\mbox{H$_2$}}
\def\Kone{\mbox{K$_1$}}     
\def\Hone{\mbox{H$_1$}}     
\def\KtwoV{\mbox{K$_{2V}$}}
\def\KtwoR{\mbox{K$_{2R}$}}
\def\KoneV{\mbox{K$_{1V}$}}
\def\KoneR{\mbox{K$_{1R}$}}
\def\HtwoV{\mbox{H$_{2V}$}}
\def\HtwoR{\mbox{H$_{2R}$}}
\def\HoneV{\mbox{H$_{1V}$}}
\def\HoneR{\mbox{H$_{1R}$}}

%RR -- Mg II h & k 
\def\hk{\mbox{h\,\&\,k}}
\def\kthree{\mbox{k$_3$}}    
\def\hthree{\mbox{h$_3$}}
\def\ktwo{\mbox{k$_2$}}
\def\htwo{\mbox{h$_2$}}
\def\kone{\mbox{k$_1$}}     
\def\hone{\mbox{h$_1$}}     
\def\ktwoV{\mbox{k$_{2V}$}}
\def\ktwoR{\mbox{k$_{2R}$}}
\def\koneV{\mbox{k$_{1V}$}}
\def\koneR{\mbox{k$_{1R}$}}
\def\htwoV{\mbox{h$_{2V}$}}
\def\htwoR{\mbox{h$_{2R}$}}
\def\honeV{\mbox{h$_{1V}$}}
\def\honeR{\mbox{h$_{1R}$}}

%%%%%%%%%%%%%%%%%%%%%%%%%%%%%%%%%%%%%%%%%%%%%%%%%%%%%%%%%% preprint header
%RR redefine @maketitle in svmult.cls to add slug on top
\ifnum\preprintheader=1     %RR ADAPT: 0 or 1 = preprintheader 
\makeatletter  %RR redefine symbol @ (trick from Pit Suetterlin)
\def\@maketitle{\newpage
\markboth{}{}%
%RR================================= Feb 27 2009 
  {\mbox{} \vspace*{-8ex} \par 
   \em \footnotesize To appear in ``Magnetic Coupling between the Interior 
       and the Atmosphere of the Sun'', eds. S.~S.~Hasan and R.~J.~Rutten, 
       Astrophysics and Space Science Proceedings, Springer-Verlag, 
       Heidelberg, Berlin, 2009.} \vspace*{-5ex} \par
%RR=================================
 \def\lastand{\ifnum\value{@inst}=2\relax
                 \unskip{} \andname\
              \else
                 \unskip \lastandname\
              \fi}%
 \def\and{\stepcounter{@auth}\relax
          \ifnum\value{@auth}=\value{@inst}%
             \lastand
          \else
             \unskip,
          \fi}%
  \raggedright
 {\Large \bfseries\boldmath
  \pretolerance=10000
  \let\\=\newline
% \@hangfrom{\@svsec}%
%%%  \@svsec
  \raggedright
  \hyphenpenalty \@M
  \interlinepenalty \@M
  \if@numart
     \chap@hangfrom{}%!!!
  \else
     \chap@hangfrom{\thechapter\thechapterend\hskip\betweenumberspace}%!!!
  \fi
  \ignorespaces
  \@title \par}\vskip .8cm
\if!\@subtitle!\else {\large \bfseries\boldmath
  \vskip -.65cm
  \pretolerance=10000
  \@subtitle \par}\vskip .8cm\fi
 \setbox0=\vbox{\setcounter{@auth}{1}\def\and{\stepcounter{@auth}}%
 \def\thanks##1{}\@author}%
 \global\value{@inst}=\value{@auth}%
 \global\value{auco}=\value{@auth}%
 \setcounter{@auth}{1}%
{\lineskip .5em
\noindent\ignorespaces
\@author\vskip.35cm}
 {\small\institutename\par}
 \ifdim\pagetotal>157\p@
     \vskip 11\p@
 \else
     \@tempdima=168\p@\advance\@tempdima by-\pagetotal
     \vskip\@tempdima
 \fi
}
\makeatother     %RR define @ back
\fi

%RR\title*{Angular-Momentum Coupling through the Tacholine}
%RD3 I checked: it was line, not cline, already in version v6
%RD3 I would not use the hyphen but don't dare to suggest this!
\title*{Angular-Momentum Coupling through the Tachocline}

\author{D. O. Gough\inst{1,2}}
%RD4 space

\authorindex{Gough, D. O.} 
%RD3 the whole book has spaces between initials because I saw
%RD3 Springer wants that.  I bet you have strong feelings on this too?
%RD3 please keep the one I put in for the index, for homogeneity.

%%\institute{Institute of Astronomy \&
%%    Department of Applied Mathematics and
%%    Theoretical Physics, University of Cambridge, UK}
%RD3 I changed the above without you noticing?  Actually, I
%RD3 prefer to do it as Springer desires:
\institute{Institute of Astronomy,  University of Cambridge, UK
    \and
    Department of Applied Mathematics and 
    Theoretical Physics, University of Cambridge, UK}

\maketitle

\setcounter{footnote}{0}  %RR Springer forgot this one (and much more)

%%%%%%%%%%%%%%%%%%%%%%%%%%%%%%%%%%%%%%%%%%%%%%%%%%%%%%%%%%%%%%%%%%%%%%%%%%%%
\begin{abstract} 
  Astronomical observation of stellar rotation suggests that at least
  the surface layers of the Sun have lost a substantial amount of the
  angular momentum that they possessed at the beginning of the
  main-sequence phase of evolution; and solar-wind observations
  indicate that magnetic coupling is still draining angular momentum
  from the Sun today.  In addition, helioseismological analysis has
  shown that the specific angular momentum at the top of the almost
  uniformly rotating radiative interior is approximately (although not 
  exactly) the same as
  the spherically averaged value at the base of the (differentially
  rotating) convection zone, suggesting that angular momentum is being
  transported through the tachocline. The mechanism by which that
  transport is taking place is not understood. Nor is there a
  consensus of opinion.  I review some of the suggestions that
  have been put forward, biassing my discussion, no doubt, according
  to my own opinions.
\end{abstract}
%%%%%%%%%%%%%%%%%%%%%%%%%%%%%%%%%%%%%%%%%%%%%%%%%%%%%%%%%%%%%%%%%%%%%%%%%%%%
%RD last sentence out?  Seems superfluous to me.
%RD but does entitle me to take out the first par below since said the same
%DOG2 Keep it in and keep the first paragraph of the original Introduction 
%DOG2 deleted, as you suggest.

%%%%%%%%%%%%%%%%%%%%%%%%%%%%%%%%%%%%%%%%%%%%%%%%%%%%%%%%%%%%%%%%%%%%%%%%%%%%
\section{Introduction}      \label{yourname-sec:introduction}
%%%%%%%%%%%%%%%%%%%%%%%%%%%%%%%%%%%%%%%%%%%%%%%%%%%%%%%%%%%%%%%%%%%%%%%%%%%%

%%RR Before we get down to the purpose of the meeting, I am here to provide a 
%%RR brief sketch of how the Sun came to be in the state of rotation that
%%RR it is today.  I shall provide only a sketch, because the evolution of
%%RR angular momentum is not well understood, and ideas are inevitably
%%RR controversial.  Consequently, quite a lot has been written and said; I
%%RR shall make no attempt to review that, but merely concentrate on those
%%RR issues that appear to be the most firmly established.

First, I recall that the Sun must have started life with much more
angular momentum than it has today.  There is considerable
circumstantial evidence to support that view, the strongest, in my
opinion being that stars are formed from interstellar gas clouds whose
typical intrinsic angular momentum per solar mass is some $10^7$ times
that of the Sun today.

Almost all of the initial angular momenta of almost all stars was lost
before the gas was even able to condense to stellar proportions.  It
is likely that during the later stages of gravitational collapse a
circumstellar disc is formed which removes angular momentum from the
star by magnetic coupling.  The process by which that occurs is a very
important arena of research, and a great deal of very interesting work
is being done to understand it.  But that is beyond the scope of this
discussion.

What I might point out, however, is that angular-momentum loss appears
to continue after the star has been established and most of the disc
has departed, as is evinced by the observed decline with age in the
rotation rates of the surfaces of stars in young clusters.  It should
also be acknowledged that even today the Sun is losing angular
momentum, via magnetic coupling to the solar wind, although at a rate
that is rather less than that suggested by the rotation rates obtained
by a simple extrapolation to the solar age of apparent
angular-momentum loss rates from young-cluster stars.

The ``simple'' extrapolation is based on the assumption that young
stars rotate almost uniformly, as does the Sun today, which may not be
realistic.  It may be that the mechanical coupling between the surface
layers and the deep interior is insufficient to transfer angular
momentum fast enough to render the surface rotation rates of young
stars direct indicators of the rotation of their interiors.  That
would explain the initial rapid slow-down, but it would also imply
that the ``initial'' angular momentum of the main-sequence Sun was
lower than average.

Was there a tachocline in those early main-sequence days?  And if so, 
was its structure similar to that of today?  I suggest that there was
a tachocline, and that its structure was similar, although in the early 
days the shear was probably stronger.

A second, very important, matter that must be considered is whether
the early Sun, and, of course, other similar young stars, harboured
substantial large-scale magnetic fields throughout their radiative
interiors; and by ``substantial'' I mean one that is intense enough to
transport angular momentum on a timescale not significantly greater
that the age of the star.  I invite you to entertain the possibility,
partly because I regard it as being most unlikely that the Sun had
lost all the field that had pervaded the interstellar gas from which
it had condensed.  Indeed, I believe that there remains a
dynamically important residual even today, simply because there
appears to be no cogent argument for maintaining the observed uniform
rotation of the radiative interior by any non-magnetic process.  I
must hastily point out that this opinion is not universally accepted,
so I shall return to it later.

Before proceeding, it is useful to state the principal properties of
the Sun that must not be ignored:  The Sun is basically in hydrostatic
balance; approximately the outer 30 per cent by radius (65\% by
volume, 2\% by mass), supporting 15 per cent of the moment of inertia,
is in a state of turbulent convection.  The rest is relatively
quiescent.  There, energy that has been generated in the
nuclear-reacting core is transported by radiative diffusion.  The
stratification of both the radiative interior and most of the
convection zone has been well established by helioseismology, as also
has the angular velocity, $\Omega$, which I now describe.

%RR fig 1
%============================================================================
\begin{figure}  
\centering
\includegraphics[width=0.6\textwidth]{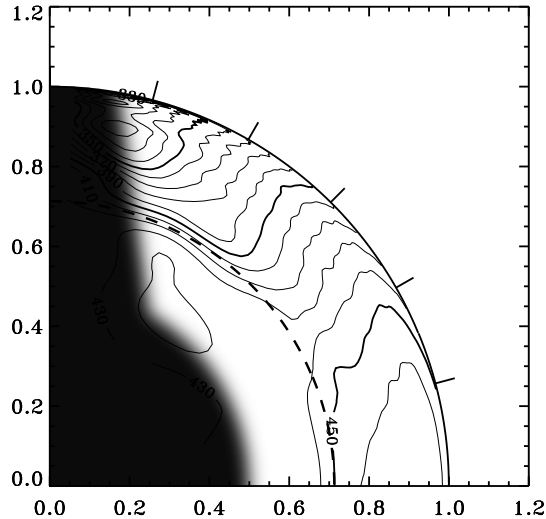}
\caption[]{ 
  Contours of constant rotation rate ($\Omega/2\pi$) in a quadrant of
  the Sun as inferred by \citet{1998ApJ...505..390S}).  The contours,
  two of which are labelled, are separated by 10~nHz.  The continuous
  quarter-circle represents the photosphere, the dashed quarter-circle
  the base of the adiabatically stratified convection zone.  The
  tachocline is clearly visible, the shear layer extending outwards
  from the radiative interior into the convection zone, the angular
  velocity increasing outwards near the equator and decreasing near
  the poles; the vertical shear vanishes at a latitude of about
  30$\deg$.  The darkened region is where reliable
  helioseismological inferences could not be drawn.
}\end{figure}
%===========================================================================

In the bulk of the convection zone $\Omega$ increases with co-latitude, from about
$2.3 \times 10^{-6}~ \rm{s}^{-1}$ (corresponding to a period of about 31
days) at the poles (where it cannot be measured accurately) to about
$2.86 \times 10^{-6}~ \rm{s}^{-1}$ (corresponding to a period of about 25.4
days) at the equator; it is roughly independent of radius, although, as 
noted by \citet{2003ESASP.517..283G}, the $\Omega$ contours are more nearly 
inclined by a constant $27\deg$ from the axis.   The
radiative interior (at least outside the energy-generating core in which 
reliable measurements have not yet been made) rotates approximately
uniformly, with angular velocity $\Omega_0 \simeq 2.67 \times 10^{-6}
~ \rm{s}^{-1}$, which is 0.93 of the equatorial angular velocity in the
convection zone (see Fig.~1).  The radiative and convective regions
are separated by a thin shear layer, known as the tachocline, which is
too thin to be reliably resolved directly by seismology (or any other
means); indirect estimates range between 2 per cent and about 8 per
%RD2 I guess you have strong feelings about ``per cent'' vs. \%? 
%DOG2 No, but I abhor percent
%RD3 there is sentence below with both per cent and % in it
cent of the solar radius, with some indication that the tachocline is
thicker near the poles than near the equator.  I should point out that
the variation in the estimates is due partly to differences in the way
the tachocline is perceived, a matter on which I shall expand later.
The base of the convection zone is likely on theoretical grounds to be
fairly accurately spherical.

%%%%%%%%%%%%%%%%%%%%%%%%%%%%%%%%%%%%%%%%%%%%%%%%%%%%%%%%%%%%%%%%%%%%%%%%%%%%
\section{Some initial remarks on tachocline dynamics}

My brief for this presentation is to discuss how angular momentum is
transported from the radiative interior to the convection zone through
the tachocline.  It is not unreasonable to suppose that the transport
has always been outwards, at least since the Sun arrived on the main
sequence.  I say this partly because the Sun is, and presumable always
has been, losing angular momentum through the photosphere to the solar
wind via magnetic stresses, and partly because most of the Sun has
been expanding slightly, and preferentially in the convection zone,
during its main-sequence evolution (the core has contracted).  It is,
of course, of interest to understand how the radiative interior has
slowed down through this process to its current, uniformly rotating,
state, which raises the question of how angular momentum is
redistributed to maintain the uniformity of rotation.  Here I shall
maintain the stance that the rotation is uniform because on timescales
of relevance the interior is rigid, held so by a large-scale
primordial magnetic field.  I shall mention points of view to the
contrary at the end of the presentation.

%RR Fig 2
%============================================================================
\begin{figure}  
\sidecaption
\includegraphics[width=0.5\textwidth]{\figspath/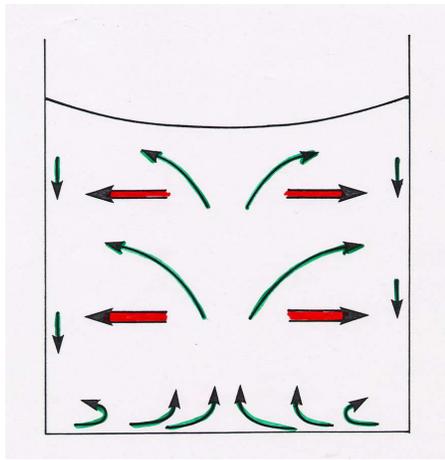}
\caption[]{
  Cartoon of a stirred cup of tea (with no handle).  The double arrows
  represent the centrifugal force excerted on the fluid, which is
  resisted by the sides of the cup inducing a radial (i.e., pointing
  away from the axis) pressure gradient which causes the surface of
  the tea to be concave.  The centrifugal force is much weaker near
  the bottom of the teacup where viscous stresses inhibit the
  rotation, leaving a (``statically'') unbalanced pressure gradient to
  drive an inward flow in a thin viscous (Ekman) boundary layer,
  denoted by the upwardly curving single arrows.  The return flow
  completing the circulation is also indicated by single arrows.
}\end{figure}
%===========================================================================

I now discuss a fundamental dynamical phenomenon which occurs
universally in rotating shearing fluids, namely that associated with
the differential rotation is a meridional circulation.  This
phenomenon was discussed by \citet{1926NW.....14..223E} to explain the
meanders of gently flowing rivers, and it is responsible for the rapid
spin-down of a stirred cup of tea. Consider the fluid in the cup
illustrated in Fig.~2. The fluid is rotating, slowly, about a vertical
axis, and the centrifugal force in the body of the fluid, denoted by
the double arrows, is resisted by the impermeable wall of the cup,
augmenting the pressure to produce an outwardly directed component to
the pressure gradient.  The higher pressure farther from the axis
supports a greater head of fluid (which is essentially in vertical
hydrostatic balance under gravity), whose upper surface is therefore
concave.  Very near the bottom of the cup, however, the rotation of
the fluid is impeded by viscous stresses against the rigid base; the
centrifugal force is unable to balance the outward component of the
pressure gradient (which evidently remains unchanged in view of the
vertical hydrostatic balance), the residual of which drives an inward
flow (denoted by the single, curved, arrows), drawing in
fluid from the side of the cup to replace it.  The inward flow is
deflected upwards into the body of the fluid, where it expands away
from the axis, and subsequently descends near the wall of the cup to
complete the circulation.  If one were to view the fluid from above,
concentrating on a portion of an outer annulus, one can liken it to a
bend in a river: the locally rotating stream produced by the bend
causes an inward flow near the bottom of the river which erodes mud,
sand and stones, transporting them from the outer bank to the inner
bank and thereby accentuating the bend, resulting finally in a
meander.

The inward flow near the base of the teacup occurs in a thin boundary
layer, of thickness $\delta$, say, whose basic structure had been
analysed by Vagn Ekman (and has subsequently been named after him) in his
doctoral thesis in 1902 for the purpose of explaining why Arctic ice
does not drift in the same direction as the prevailing wind.  It
explains also why a stirred cup of tea slows down faster than the
characteristic viscous diffusion time $\tau_{\rm{d}}=R^2/\nu$, where
$R$ is the radius of the teacup and $\nu$ is the viscous diffusion
coefficient: the angular momentum in the tea is transferred to the cup
in the boundary layer on the shorter timescale $\delta ^2/\nu$, and
the angular velocity in the essentially inviscid flow elsewhere
declines because the angular-momentum-conserving fluid is drifting
away from the axis of rotation (it can be shown that the flow near the
outer wall occurs in a layer considerably thicker than $\delta$, so
angular-momentum transfer there can be ignored).  When all this is put
together, the final outcome, as in so many other cases of fluid flow
with thin diffusive (boundary) layers, is that global evolution occurs
on a timescale which is the geometric mean of the characteristic
dynamical and the large-scale diffusive times.  In the case of a cup
of tea, $\tau_{\rm{d}} \simeq 1/2$~hour, and if one has stirred at a
rate of two per second, the dynamical time is 1/2~second; therefore
the tea slows down in about 1/2~minute.  Note that that
spin-down time is also the characteristic circulation timescale.
%RD2 I replaced the \fracs by ratios - fracs get too small in in-line math
%DOG2 Fine

%RR Fig 3
%============================================================================
\begin{figure}  
\sidecaption
\includegraphics[width=0.5\textwidth]{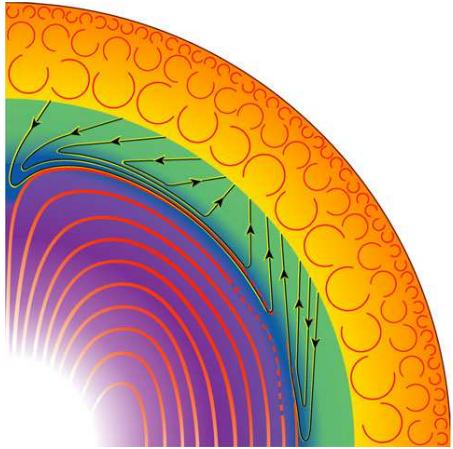}
\caption[]{
  Quadrant of the Sun indicating the gyroscopically driven meridional
  flow (lines with arrows) in the tachocline (from
  \cite{1998Natur.394..755G}). The width of the tachocline has been
  artificially exaggerated to render the flow visible.  Circular arcs
  represent convective eddies, and the broad lines beneath the
  tachocline represent lines of magnetic field -- they are dashed near
  the upwelling region, where their geometry is even more uncertain
  than elsewhere and where they are probably drawn into the convection
  zone, providing a source of Maxwell stress that may locally quench
  the shear.  
}
\end{figure}
%===========================================================================

The solar tachocline is broadly similar.  The rigid radiative
interior, which is very strongly stably stratified (the characteristic
buoyancy restoring timescale is about an hour -- the characteristic
period of grave g modes -- which is very much shorter than even the
rotation period), acts like the rigid impermeable base of the teacup,
above which is fluid which is caused to rotate differentially by the
anisotropic Reynolds stresses, and associated Maxwell stresses, in the
convection zone.  Even the simplest dynamical description of the
boundary layer (tachocline) that separates the two regions is
considerably more complicated than an Ekman boundary layer 
(\cite{1998Natur.394..755G}); 
the basic dynamics of the meridional circulation in
the convection zone is also somewhat different, being maintained by
``gyroscopic pumping'' by the anisotropic turbulent stresses rather than
a spinning down like a cup of tea that is no longer being stirred.  Near
the equator, where the fluid in the convection zone is rotating faster
than the ``rigid'' interior, there is a (poleward) essentially
horizontal flow in the tachocline towards the axis of rotation;
contrarily, near the poles, where the convection zone rotates more
slowly than the interior, the tachocline flow is equatorward.  In both
regions the motion draws fluid in from the convection zone (or, if one
prefers, is pushed aside by the fluid descending from the convection
zone from whence it has been pumped by the rotating anisotropic
turbulent stresses).  The opposing flows in the tachocline meet
somewhere at mid latitudes, and, being unable to penetrate
significantly into the radiative zone below, they are forced to rise
back into the zone above, as in a cup of tea.  The flow is illustrated
in Fig.~3, where lines of interior magnetic field are also depicted.

%%%%%%%%%%%%%%%%%%%%%%%%%%%%%%%%%%%%%%%%%%%%%%%%%%%%%%%%%%%%%%%%%%%%%%%%%%%%
\section{The early history of dynamical tachocline models}

The existence of the tachocline was first recognized theoretically by
\citet{1972NASSP.300...61S}
in his study of the spin-down of the Sun.  It attracted little
attention at the time, and was hardly discussed until the existence of
the tachocline was established seismologically 
(\cite{1987ApJ...314L..21B};
 \cite{1988ESASP.286..149C};
 \cite{1989ApJ...343..526B};
 \cite{1989ApJ...337L..53D}).
A prime motivation for the early consequent studies was
to explain how the radiative interior of the Sun could rotate
uniformly in the face of the non-uniform stresses imposed by the
differentially rotating convection zone, although many authors make a
different emphasis, addressing why the transition between the
differentially rotating convection zone and a putative uniformly
rotating interior is so thin.  The conclusion is that a thin
tachocline is indeed plausible, perhaps even inevitable.

The first dynamical discussion of the matter was provided by 
\citet{1992A&A...265..106S},
who proposed that turbulence in the stably stratified
tachocline was essentially layerwise two-dimensional, having
negligible vertical motion and being horizontally isotropic so that it
acted like a powerful (constant) viscosity on horizontal spheres. The
turbulence appeared to be generated by the latitudal rotational shear
itself, and was presumed to be sufficiently intense actually to quench
at depth the shear that produces it.  Subject to those assumptions a tachocline was
elegantly constructed so as to reproduce the helioseismological
findings.  

One of the obvious deficiencies of the Spiegel-Zahn model is that it
 predicts the wrong angular velocity $\Omega_0$ for the uniformly rotating 
core: $0.90\,\Omega{_{\rm{eq}}}$ rather than the observed value of 
$0.93\,\Omega{_{\rm{eq}}}$, where $\Omega{_{\rm{eq}}}$ is the equatorial 
angular velocity in the convection zone.  This quantity had already been 
discussed by 
\citet{1985shpp.rept..183G},
who pointed out that its value would have been
$0.96\,\Omega{_{\rm{eq}}}$ had the local shear stress--rate-of-strain
relation at the base of the convection been independent of latitude,
as would be the case with a thin tachocline exerting a shear stress
produced by a uniform isotropic viscosity and (artificially)
supporting no meridional flow.  It is an important diagnostic of
tachocline models, partly because it is an overall measure of the
differential angular-momentum transport and partly because its
latitudinal average can be measured quite precisely by seismology.
Indeed, in those days it appeared that the spherically averaged
angular velocity in the radiative interior might drop below even
$0.90\,\Omega{_{\rm{eq}}}$ 
(\cite{1984Natur.310...22D};  % Duvall et al., 1984;
 \cite{1984AdSpR...4...85G}),  % Gough, 1984), 
which might actually be the case in the core 
(\cite{1995Natur.376..669E};  % Elsworth et al., 1995; 
 \cite{1997MNRAS.292..243B}). % Basu et al., 1997).  
The reduction of $\Omega_0$ in the Spiegel-Zahn model appears to
result principally from the latitudinal transport of angular momentum
by the shear-turbulent viscous stresses in the tachocline acting upon
the meridional circulation, which behave differently from the Reynolds
stresses in the convection zone resulting from three-dimensional
turbulence driven by buoyancy.

Spiegel \& Zahn's assumptions were challenged some time
later (\cite{1998Natur.394..755G}), principally because rotating
layerwise two-dimensional turbulence in nature does not act in such a
way as to quench shear.  Nor, in general, does any other known purely
fluid-dynamical (non-magnetic) motion, such as the dissipation of
gravity waves generated near the base of the convection zone, which
\citet{1997ApJ...475L.143K} and \citet{1997A&A...322..320Z} 
had erroneously 
(e.g., \cite{1997Natur.388..324G})  % Gough, 1997)
%RD2 yes, I confused 1977 and 1997 (intact retinae but plastic inserts)
argued.  It was therefore concluded that only if the radiative
interior were held rigid by a sufficiently intense magnetic field
could its angular velocity be uniform.  The flow descending into the
tachocline at low and high latitudes (see Fig.~3) advects the
outwardly diffusing magnetic field to maintain a field-confining
balance which plays a role in determining the thickness of the
tachocline.  At mid latitudes, where the rotational shear is small,
the ascending return flow is likely to drag field into the convection
zone.  There the dynamics is difficult to analyse, which is why the
field lines in Fig.~3 have been drawn dashed.  In their admittedly
grossly oversimplified model, Gough \& McIntyre assumed the field to
be basically dipolar (a reasonable assumption, because the dipole
component of a complicated primordial field decays the most slowly),
purely poloidal (not so reasonable because such fields are not stable
(\cite{1973MNRAS.163...77M});
a numerical simulation of field decay in an
idealized star by \citet{2004Natur.431..819B} resulted in a dipole
field with a toroidal component), and with, for simplicity, its axis
aligned with the axis of rotation. Although those assumptions are
bound not to be satisfied exactly, it is unlikely that any plausible
deviation from them would invalidate the basic underlying
arguments. In both the Spiegel-Zahn and the Gough-McIntyre models,
angular momentum is advected through the bulk of the tachocline by the
meridional circulation.  It appears to be the case that the overall
strength of that flow is insensitive to the details of the tachocline
boundary-layer structure, as is perhaps the case also of the cup of
tea (recall that the time to spin down a cup of tea is the geometric
mean of two global times that take no explicit account of the Ekman
boundary-layer structure); the ventilation timescale (quoted by Gough
and McIntyre but not explicitly by Spiegel and Zahn) of the meridional
circulation in the solar tachocline is of order $10^6~$y.
%RD3 I would use year here

It has often been mentioned that \citet{1997AN....318..273R} had
already pointed out that the radiative interior could be held rigid by 
a horizontal magnetic field of sufficient intensity, and that the shearing 
interface could be thin if the vertical variation of the field were 
somehow maintained.  However, no cogent dynamical argument addressing how 
the field could be confined was offered.  Instead, it was merely stated 
that the field would not diffuse into the convection zone because the 
(eddy) diffusivity there is so high;  no explicit account was taken of the 
dynamical implications of the necessarily consequent high diffusion rate.

In the model by Gough \& McIntyre, an additional angular-momentum
conduit is provided by the magnetic field that is presumed to
penetrate the tachocline at mid latitudes.  In the absence of
instability (an unrealistic circumstance, but one to consider as a
benchmark), any (imposed) vertical rotational shear would stretch the
field into a gyre, creating an increasing opposing torque until
diffusion quenches its growth.  Perhaps that torque quenches the shear
instead, leading to a tachocline with essentially no shear at all
(with apologies for the oxymoron) in the region (in each hemisphere)
of penetrating field. T. Sekii and I are currently seeking
seismological evidence for such a shear-free region, which, according
to current seismological inference (see Fig.~1), must be located near
latitudes $\pm 30\deg$.

The $10^6$-year tachocline ventilation time is very much shorter than
the $10^{10}$-year evolution timescale of the Sun.  Therefore, on that
timescale the net transport of angular momentum across the tachocline
essential vanishes, unless the Sun is undergoing a deeply seated
torsional oscillation (for which there is actually some
helioseismological evidence, provided by \citet{2000Sci...287.2456H},
who report a 1.3-y oscillation in $\Omega$ near the equator
immediately above and below the tachocline), in which case it is the
time-averaged angular momentum transport that vanishes.  That
constraint, coupled with the (model-dependent) latitudinal variation
of the angular-momentum flux, determines the equilibrium (average)
angular velocity $\Omega_0$ of the radiative interior.

Note that I have in mind a state of rotation which is essentially
steady (with perhaps a gentle superposed oscillation) over the
ventilation timescale; on the much longer timescale of the
main-sequence evolution of the Sun, a minute shear would transport
angular momentum enough to maintain the angular-velocity balance
between the braking convection zone and the radiative interior.

I should point out that the confinement of the magnetic field by the
generally downwelling meridional flow in the radiative interior, except
in the presumably small upwelling zone at mid latitudes, is an
essential feature of the dynamics proposed by Gough \&
McIntyre. Without it, magnetic field would thread the tachocline into
the convection zone where it would, in the mean, adopt (much of) the 
differential rotation of that zone and thereby exert a torque on the
radiative region below, causing it to rotate differentially too.  Some
numerical simulations have provided examples of this general process
(e.g., 
\cite{1999ApJ...519..911M}; 
\cite{2002MNRAS.329....1G};
\cite{2006A&A...457..665B}).

An essential property of the dynamics, therefore, is that the
downwelling velocity is strong enough to dominate the outward
diffusion of the field, except of course at the very bottom of the
tachocline where there is magnetic boundary layer -- the magnetopause
-- of thickness $\delta$, which is much thinner than the thickness
$\Delta$ of the essentially diffusion-free body of the tachocline.  It
goes without saying that in order to model the downwelling from the
convection zone the fluid must be permitted to flow from the
convection zone into the tachocline; any theoretical model separating
the two regions by an artificial impermeable boundary is bound not to
provide a faithful representation of the true dynamics.

%RR Fig 4
%============================================================================
\begin{figure}  
\centering
\includegraphics[width=\textwidth]{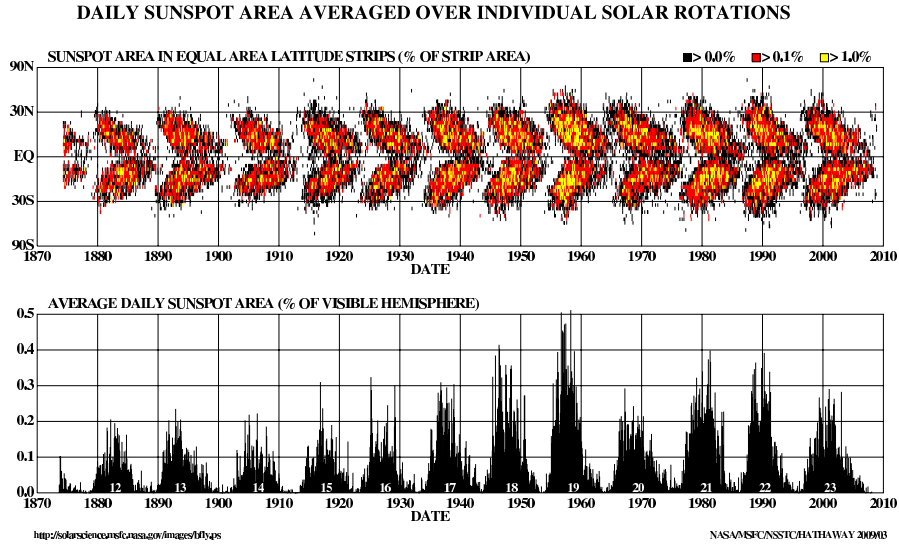}
\caption[]{
  Sunspot area (averaged over a solar rotation) plotted against time.
  The lower panel is a histogram of the total area; the upper panel
  indicates the latitudes of the spots (courtesy D.~Hathaway, NASA
  Marshall Space Flight Center).
}\end{figure}
%===========================================================================

%%%%%%%%%%%%%%%%%%%%%%%%%%%%%%%%%%%%%%%%%%%%%%%%%%%%%%%%%%%%%%%%%%%%%%%%%%%%
\section{Some consequences of tachocline circulation}

It can hardly pass unnoticed that the latitude at which magnetic field
might penetrate the tachocline is the same as that at which sunspots
first appear at the start of a new cycle (see Fig.~4).  Perhaps it is
the radiative interior of the Sun that supplies the convection zone
with a seed field with which to generate the magnetic cycle.  I shall 
leave that remark undeveloped, one on which to ponder.
Evidently magnetoydrodynamical processes, commonly called dynamo
action in this context, process the field to produce its cyclic
behaviour; but those processes may not sustain the field as in a true
dynamo, instead declining with the decaying primordial field.

%RR Fig 5
%============================================================================
\begin{figure}  
\centering
\includegraphics[width=0.85\textwidth]{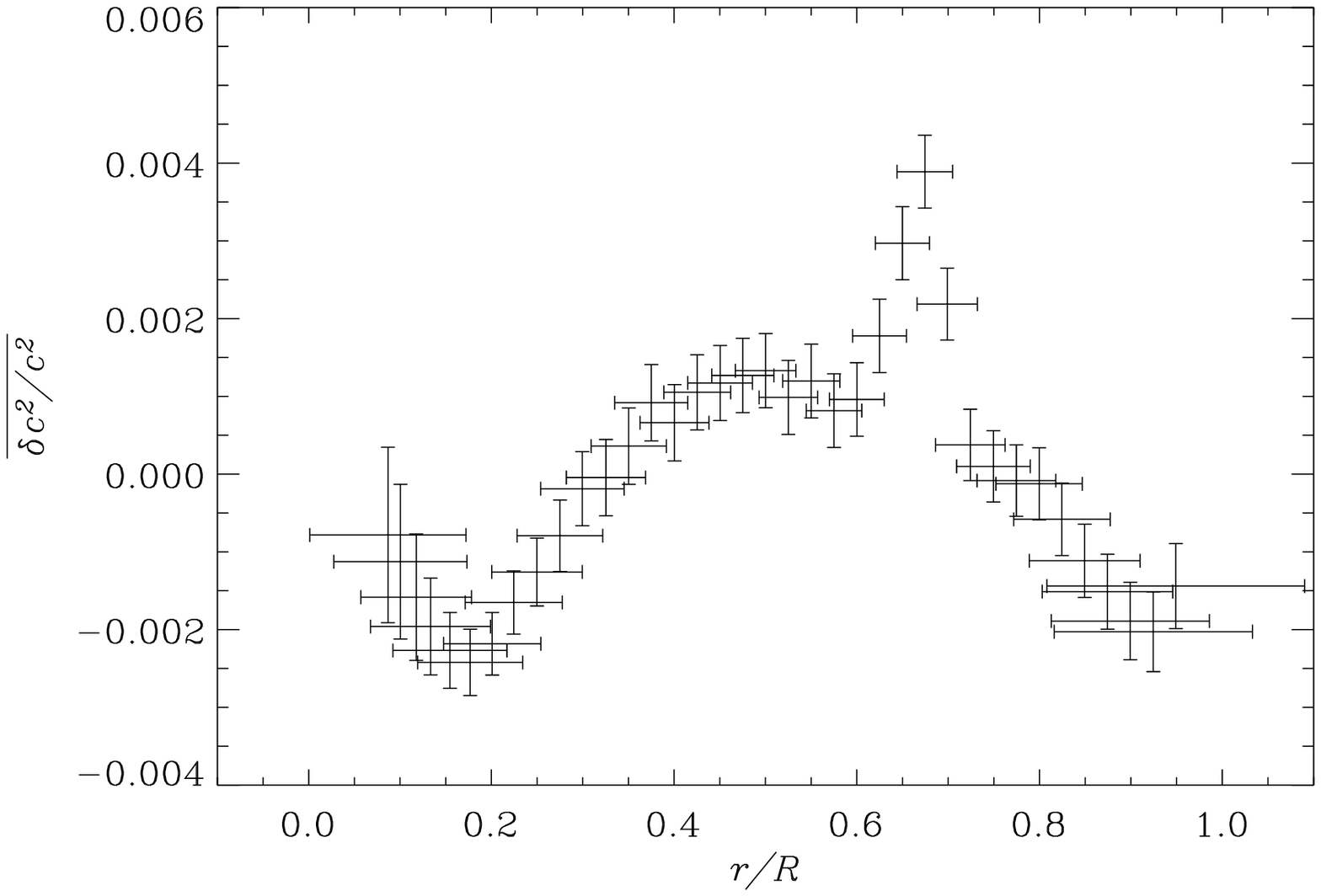}
\caption[]{
  Localized averages $\overline{\delta c^2/c^2}$ of the relative
  differences of the (spherically averaged) squares of the sound
  speeds in the Sun and in a standard theoretical solar model (Model S
  of \cite{1996Sci...272.1286C}), plotted against the location of the
  centres of the overaging kernels.  The horizontal bars indicate the
  extent of the averaging (approximately full width at half maximum of
  the almost Gaussian averaging kernels), the vertical bars the
  standard errors (which are correlated with one another) arising from
  the formal uncertainties in the seismic frequency data, which are
  presumed to be independent of one another.  The prominent sharp
  anomaly is in the region occupied by the tachocline beneath the
  convection zone, the base of which is situated at radius $r =
  0.713\,R$, where here $R$ is the radius of the Sun
%RD2 added _odot
%DOG2 _odot removed for consistency with figure --  see e-mail
%RD3 apology, I never noted the R in the figure
  (\cite{1991ApJ...378..413C}).
}\end{figure}
%===========================================================================

One process of which we can be sure is that the meridional
circulation, driven by gyroscopic pumping in the convection zone,
ventilates the tachocline and tends to mix any heavy elements that may
have settled under gravity back into the convection zone. Since the 
ventilation time is very much less than the age of the Sun,
that mixing is essentially complete in the Gough-McIntyre scenario; in
the Spiegel-Zahn scenario the rapid horizontal diffusion opposes the
vertical transport, augmenting the mixing time to a substantial
fraction of the solar age 
(\cite{1992A&A...253..173C})
and rendering it
plausible that mixing is incomplete.  In either case the outcome is a
local reduction of the mean molecular mass of the tachocline material 
below that predicted by
standard solar models, and a consequent augmentation of the sound
speed, producing what has been called the tachocline anomaly (see Fig.~5).
Calibration of this anomaly seismologically is the most precise way to
determine the thickness of the tachocline because it uses the
multiplet frequencies (mean over azimuthal order at given principal
order and degree) of the seismic modes, rather than the less-well
determined rotational splitting.  However, the outcome depends on the
premises upon which the theoretical reference solar model was built,
and on the presumed evolution of the thickness of the tachocline.  It
may therefore not provide the most accurate determination.

It should be pointed out that the molecular-mass gradient (maybe a near 
discontinuity) created at the base of the tachocline is hydrostatically  
very stable: that is to say, that it cannot be significantly upset by the 
$10^6$-year meridional flow or the magnetic field that that flow confines 
(a basic point that was missed by Gough \& McIntyre when they suggested the
possibility of a ``polar pit'' in the tachocline which might permit mixing 
of lithium to depths at which nuclear transmutations could occur today). 
Therefore the base of the tachocline is likely to be quite precisely 
spherical (actually, almost coincident with an equipotential surface).

It is appropriate here to point out that the tachocline calibration
yields the thickness of only the layer in the radiative region that is
homogenized with the convection zone.  That is precisely the layer
which Spiegel and Zahn named the tachocline.  Spiegel and Zahn, and
Gough \& McIntyre after them, assumed for simplicity that the natural
stresses in the convection zone are so strong that the tachocline does
not react back on the angular velocity in the convection zone by an
appreciable amount, so that essentially all the rotational shear
resides in the radiative layer.  Helioseismological evidence suggests
that that is not exactly right (see Fig.~1), and that the shear has
spread somewhat into the convection zone above, especially at high
latitudes where vortex stretching is the greatest.  There the dynamics
is quite different; so it is perhaps prudent to reserve the term
``tachocline'' for only the radiative boundary layer, respecting
Spiegel \& Zahn's precedent notwithstanding its etymological origins.
A straightforward calibration of the sound-speed anomaly can then
yield a tachocline thickness $\Delta$ of about $0.02\,{\rm{R}}_\odot$,
%RD2 R is a physics param and should be in math italics
%DOG2  May be, but $\rm{R}_\odot$ is a recognized unit, which is what I mean.
%DOG2  Had I used  $R_\odot}$, which would have been possible, the sentence 
%DOG2  here (and elsewhere) would have needed rephrasing to accommodate the
%DOG2  different meaning.
%RD3 OK, as you like. I don't see R_sun as unit but I now see what you mean
assuming complete mixing (\cite{1999ApJ...516..475E}).  Incomplete
mixing would yield a larger value.  (At this point it is worth
mentioning that complete mixing down to some level is not consistent
with a recent seismological analysis by Christensen-Dalsgaard and
Gough (unpublished), but nor is a simple incompletely mixed layer
residing entirely beneath the adiabatically stratified convection
zone.)  Calibrations by fitting parametrized functions to the
seismologically inferred shear layer near the base of the convection
zone 
(e.g.,
\cite{1996ApJ...469L..61K};
\cite{1999ApJ...526..523C};
\cite{1999A&A...344..696C}), 
however, depend, of course, on the fitting function adopted.  They
have yielded values of $\Delta$ ranging from about $0.03\,{\rm{R}}_\odot$,  
(near the equator) to $0.08\,{\rm{R}}_\odot$,(near the
poles).  \citet{2003ApJ...585..553B}, for example, have reported that
the surface defined by the mid-point of the transition is prolate; one
can infer from their results, accepting the quoted uncertainties
literally, that the base of the tachocline, defined by the shear,
could perhaps be prolate too (with a pole-equator radius difference of
about $0.01\,{\rm{R}}_\odot$), but in the light of my earlier dynamical
remarks on the matter I advise taking such a conclusion with a pinch
of salt. 
%No convincing seismological evidence for solar-cycle
%variation of tachocline structure has been reported.
%DOG
There is some seismic evidence for solar-cycle variation
of tachocline structure (e.g., 
\cite{1989ApJ...347..540D}; % Dziembowski and Goode, 1989; 
\cite{2008ApJ...686.1349B}), % Baldner and Basu, 2008),
but its nature is undetermined.

%%%%%%%%%%%%%%%%%%%%%%%%%%%%%%%%%%%%%%%%%%%%%%%%%%%%%%%%%%%%%%%%%%%%%%%%%%%%
\section{The quest for a tachocline simulation}

There have been a series of attempts to address whether a
magnetic-field-confining tachocline can actually be formed and maintained,
using numerical simulations of different degrees of sophistication.
Perhaps the most ambitious is a dynamical anelastic computation
of \citet{2006A&A...457..665B}, who modelled the outer half (by radius) of
the radiative zone of a rotating Sun-like star harbouring a
large-scale magnetic field.  The inner and outer boundaries, at radii
$0.35\,{\rm{R}}_\odot$ and $0.72\,{\rm{R}}_\odot$, were presumed to
be impermeable but otherwise stress-free, maintaining a constant
radial component of the entropy gradient; the magnetic field was
matched smoothly onto potential fields outside the domain of
computation.  Of course, realistic diffusion coefficients could not
be adopted; they had to be taken many orders of magnitude greater, but
Brun and Zahn wisely maintained the correct ordering of their
magnitudes, which they believed captured the essence of the real
problem. They chose constant values, formally augmenting the thermal 
diffusivity $\kappa$ (in the tachocline) by a factor $6 \times 10^5$,
the magnetic diffusivity $\eta$ by $2 \times 10^8$ and the viscosity 
by $3 \times 10^8$ (taking the diffusion coefficients in the tachocline 
to be those quoted by 
\cite{2007sota.conf....3G}).

Brun \& Zahn carried out their computations from a variety of initial
states, the most promising (for anyone interested in seeing a positive
outcome, one might think) being one with a field well confined within the
radiative interior.  In all cases the field diffused through any
semblance of a tachocline into the convection zone and took up its
differential rotation, imparting it to the radiative interior
beneath.  In most of the simulations a potential magnetopause would
have occupied the entire tachocline ($\delta \simeq \Delta$, rather
than $\delta \ll \Delta$ as found by Gough \& McIntyre).  A
simulation was run with $\delta \simeq \Delta/3$; the differences in
the result were barely noticeable.  Brun and Zahn concluded that other
processes must therefore be invoked to explain the tachocline
structure, such as gravity-wave transport of the kind discussed by
%RD Talon and Charbonnel 
%RD wrong order wrt bib file
(\cite{2005Sci...309.2189C}; see also \cite{2007AIPC..948...15C}) or a
rigidly generated oscillating dynamo field of the kind considered by
Forg\'{a}cs-Dajka \& Petrovay 
(2001, 2002); 
\nocite{2001SoPh..203..195F} 
\nocite{2002A&A...389..629F}
I shall return to those matters later.

%RR Fig 6
%============================================================================
\begin{figure}  
\centering
\includegraphics[width=0.48\textwidth]{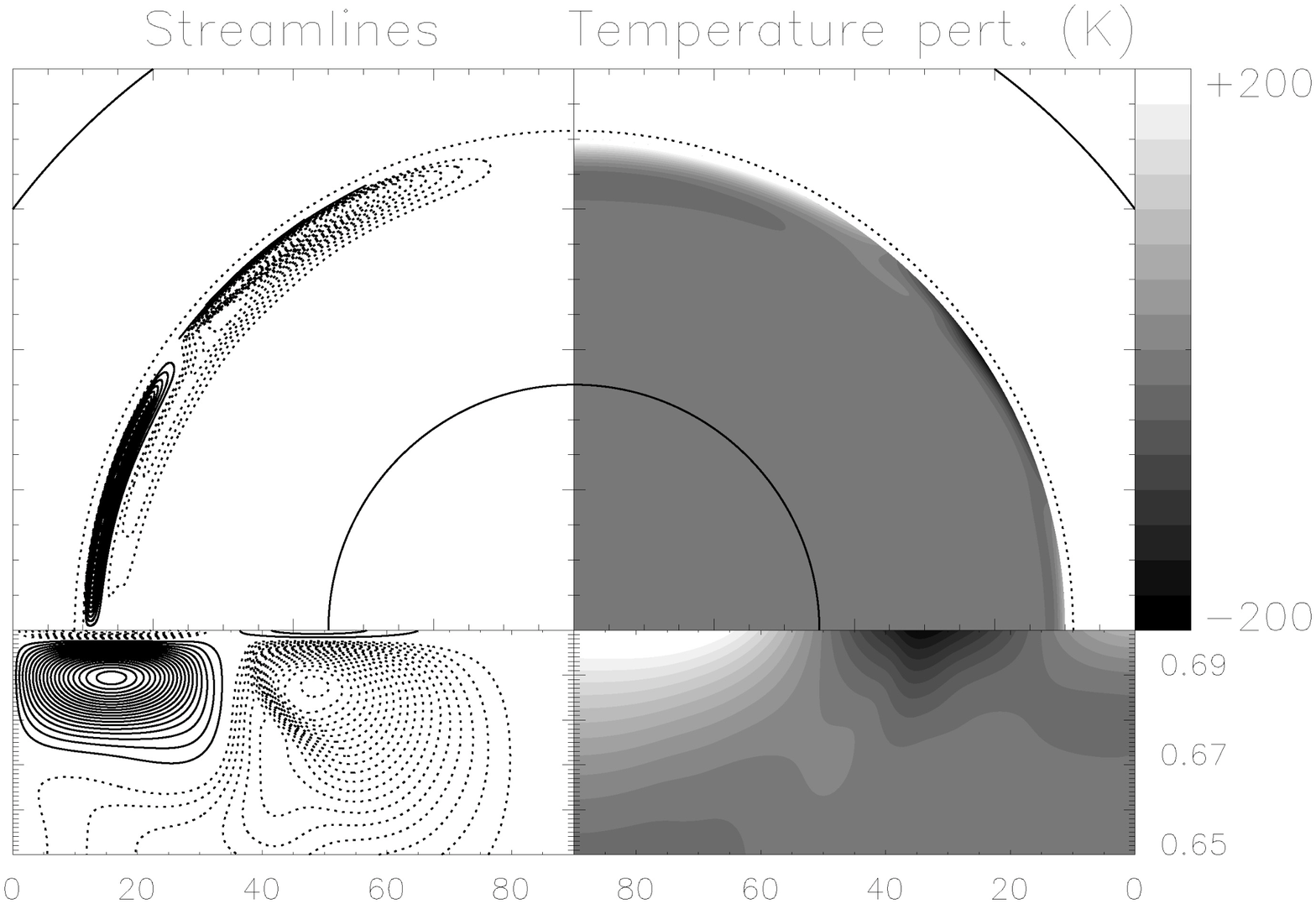}
\includegraphics[width=0.46\textwidth]{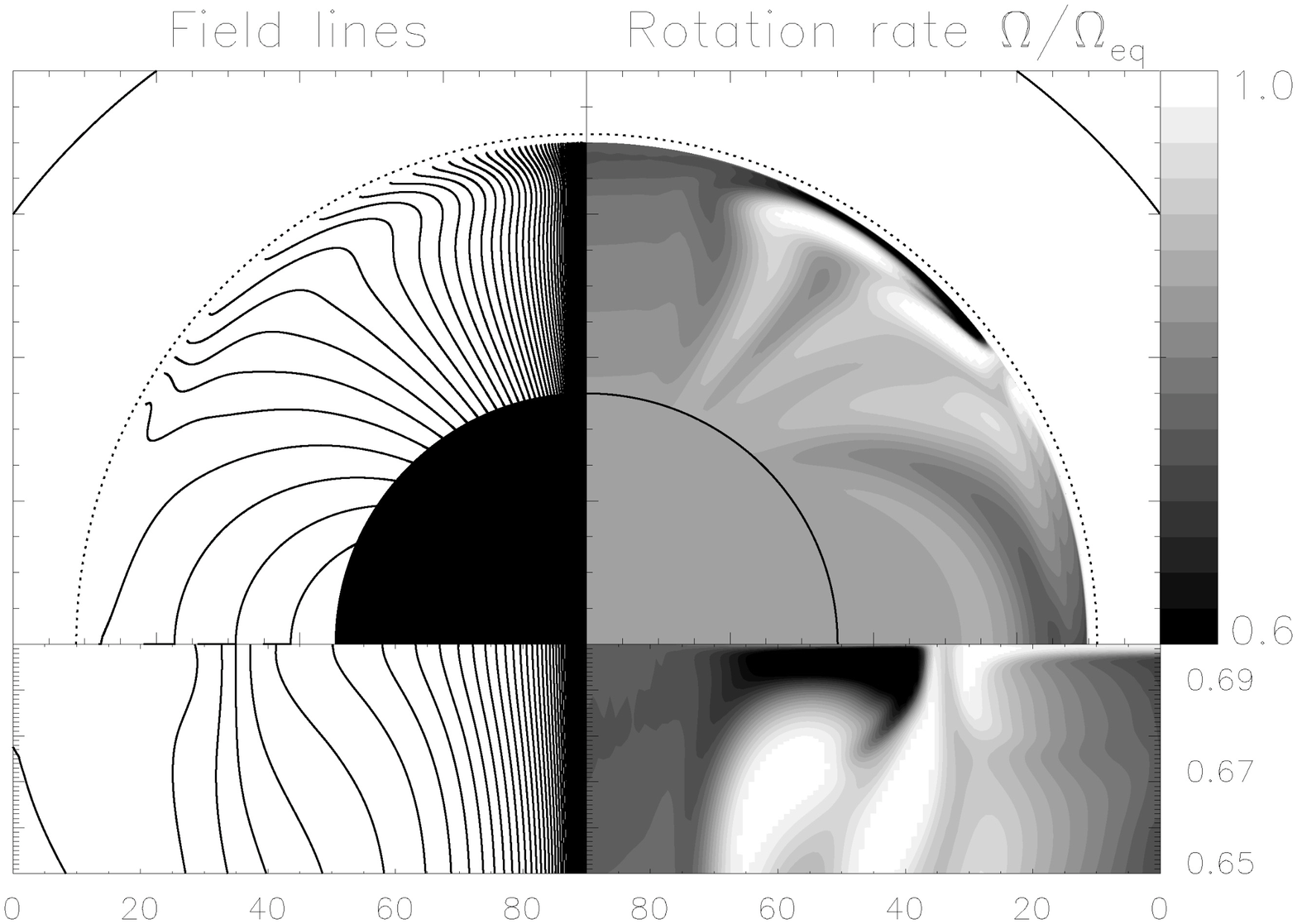}
\caption[]{
%DOG I've extended the first sentence.
  Tachocline model by \citet{2008MNRAS.391.1239G}, with imposed
  vertical velocity through the upper boundary as in Fig.~7.  In the
  quadrants are indicated (on the left) streamlines (left) -- solid for
  clockwise circulation, dashed for anticlockwise -- and magnetic
  field lines (right), and (on the right) greyscale representations of
  temperature perturbations from the spherically symmetric background
  state (left) and the angular velocity $\Omega$ (right) measured in units of the
  equatorial angular velocity $\Omega{_{\rm{eq}}}$ of the convection
  zone.  Beneath each quadrant is an enlargement of the tachocline
  region, plotted against latitude and radius.
}\end{figure}
%===========================================================================

%RR Fig 7
%============================================================================
\begin{figure}  
\sidecaption
\includegraphics[width=0.5\textwidth]{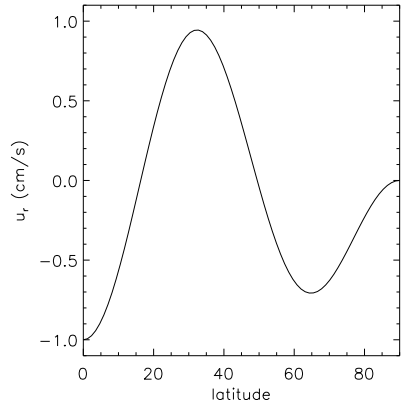}
\caption[]{
  Radial velocity component imposed by \citet{2008MNRAS.391.1239G} at
  the base of the putative convection zone, plotted as a function of
  latitude.
}\end{figure}
%===========================================================================

A different approach has been taken by Garaud and her colleagues -- the 
discussion by 
\citet{2007sota.conf..147G} % Garaud (2007) 
puts it into context -- who carried out steady-state axisymmetric
computations of a model of the radiative zone, the temporal and
geometrical simplifications being made with the intention of rendering
it possible to reduce the diffusion coefficients well below those
truly attainable in resolved three-dimensional time-dependent
simulations, and, it is hoped, to values even low enough for the
interior magnetic field to be contained by the tachocline flow.  She
presumed the convection zone to be magnetically infinitely diffusive,
and in her first paper 
(\cite{2002MNRAS.329....1G})  % (Garaud, 2002) 
adopted an impermeable but
otherwise stress-free boundary between the convection zone and the
radiative interior, as did \citet{2006A&A...457..665B} after her.  She
too always found the magnetic field to penetrate the tachocline,
particularly in the polar regions, and so impart a differential
rotation on the radiative interior.  But with her realization of the
importance of the downwelling from the convection zone at low and high
latitudes driven by the gyroscopic pumping
(\cite{2008ApJ...674..498G}), and the development of a new numerical
procedure, she succeeded in collaboration with her brother
(\cite{2008MNRAS.391.1239G}) to compute models of the outer part of
the radiative zone ($0.35\,{\rm{R}}_{\odot}
\leqslant \it{r} \leqslant \it{r}{_{\rm{c}}}$ with $r{_{\rm{c}}} =
0.7\,{\rm{R}}_\odot$ being the presumed base of the convection zone) 
in which fluid with yet lesser diffusion is made to descend from and
ascend into the convection zone in a prescribed way.  

\citet{2008MNRAS.391.1239G} demonstrated that if the upper boundary
were presumed to be impermeable, the vertical flow could never be
strong enough to attain a magnetic Reynolds number high enough to
contain the field, no matter how low the diffusion coefficients might
be, thereby explaining why all previous attempts to obtain a plausible
numerical model of the tachocline had failed.  When motion through the
upper boundary was introduced, however, a dynamical balance between
advection and diffusion of magnetic field qualitatively similar to
that described by
\citet{1998Natur.394..755G}, with angular momentum transported
principally by advection in the upper part of the tachocline and by
Maxwell stresses in the lower part, could in some situations be
achieved, although the solutions also revealed a number of previously
unrecognized subtelties.  Moreover, the configuration adopted by the
flow and the field is not unlike that found in the recent study by
\citet{2007AIPC..948..303W} which extended the analysis of the
Gough-McIntyre model to the vicinity of the poles.

As in all numerical studies to date it was necessary for Garaud and Garaud 
to augment the diffusivities by large factors, which they did in such a way 
as to maintain the hierarchy of characteristic boundary-layer thicknesses 
that might be expected to be encountered: the magnetic diffusivity $\eta$ 
was augmented by a factor $\tilde{f}$, and the kinematic viscosity $\nu$ and 
radiative thermal diffusivity $\kappa$ were adjusted accordingly to yield a 
Prandtl number $\nu/\kappa$ that is 10 times solar and a magnetic Prandtl 
%DOG   Fig. 6 referred to in next line
number $\nu/\eta$ that is 0.1 solar.  Figure~6 shows the solution for
the least diffusive case that Garaud \& Garaud were able to compute
with an upper permeable boundary, having $\tilde{f} = 8 \times 10^9$.
(It should be pointed out that this value is actually more than 10
times higher that what was formally employed by
\citet{2006A&A...457..665B}, 
but on the other hand the spatial resolution of the numerical scheme
that was employed was more than 10 times finer, sufficient to resolve
the boundary layers that were anticipated.  Garaud \& Garaud did
reproduce the Brun-Zahn coefficients with an impermeable boundary.)
Illustrated are streamlines, magnetic field lines, and greyscale
representations of temperature perturbations and the angular velocity
in quadrants of the Sun, underneath which are enlargements of the
tachocline region ($r>0.65\,{\rm{R}}_\odot$) plotted as functions of
latitude and radius, $r$.  It is evident that a tachocline has been
established, and that, for the first time in a global simulation, the
large-scale behaviour of the differential rotation of the convection
zone has not been imprinted at great depths in the radiative interior.
There is some spatial variation in the interior angular velocity,
however, on a relatively small scale which is not evident in Fig.~1.
Garaud \& Garaud suggest that the averaging kernels for Fig.~1 are
broad enough not to have resolved the structure, and that in any case
the features are probably much stronger than is likely to exist in the
Sun because it had been necessary to augment the magnitudes of both
the imposed flow velocities and the magnetic field in order to produce
a magnetic Reynolds number that exceeds unity with the enormously
increased diffusivities that it had been necessary to adopt.

In the face of these encouraging beginnings there remain some
outstanding issues.  The most blatant is the angular velocity
$\Omega_0$ of the inert core, which is in torque-free equilibrium with
the outer part of the radiative zone.  Its value is
$0.875\,\Omega{_{\rm{eq}}}$, whereas seismological analysis of the Sun, 
as I have already pointed out, 
yields $0.93\,\Omega{_{\rm{eq}}}$, $\Omega{_{\rm{eq}}}$ being the
equatorial angular velocity of the convection zone.  The discrepancy
is unexplained, although \citet{2008arXiv0811.2550G} are making some
progress towards rectifying that.  The discrepancy may be merely a
product of not having achieved a low-enough value of the diffusivity
augmentation factor $\tilde{f}$.

Another obvious deficiency of the model is that the vertical flow through 
the base of the convection zone has been imposed artificially.  A 
significant advance would be to couple the tachocline to a simple yet not 
too unrealistic model of the convection zone, with, say, a representation 
of the effect of the anisotropic Reynolds stresses on the large-scale flow 
to generate the gyroscopic pumping.  I eagerly await the outcome of such a 
calculation.

%%%%%%%%%%%%%%%%%%%%%%%%%%%%%%%%%%%%%%%%%%%%%%%%%%%%%%%%%%%%%%%%%%%%%%%%%%%%
\section{Closing Remarks}

I conclude with three brief remarks concerning alternative views of
the overall dynamics.  I trust it is evident, even if I have not given
it as much emphasis as others might, that the general picture that I
have advocated is far from being universally accepted.  As I pointed
out in the previous section, \citet{2006A&A...457..665B}, for example,
believe that the principal model I have discussed lacks some further
essential ingredient.  I shall do little more than simply mention
them.

My first point concerns the angular-momentum transport by the
layerwise two-dimensional turbulence envisaged by
\citet{1992A&A...265..106S} in their tachocline model.  As
\citet{1998Natur.394..755G} pointed out, such motion is not expected
to lead to diffusive mixing of angular velocity -- indeed it is not
even everywhere diffusive of anything -- but in places transports
angular momentum in a wave-like manner -- it is what some people have
called antidiffusive, with a tendency to mix potential vorticity
(\cite{2001MNRAS.324...68G}; see also \cite{1968JFM....32..437G}).
\citet{2003safd.book..111M} provides a more extensive discussion,
citing a critical analysis by \citet{1991JAtS...48..651H} showing that at
least in the rectilinear case straightforward local momentum mixing by
turbulence is not possible without accompanying wave transport.
However, there is a more recent, and rather different, numerical
simulation by
\citet{2003ApJ...586..663M} which appears to lead Miesch to conclude
the contrary, and which therefore demands comment.  Miesch considered
a spherical shell of fluid between impenetrable, isothermal,
horizontal, stress-free boundaries, initially rotating uniformly with
angular velocity $\Omega_0$.  He supplied the vertical component of
the momentum equation with an axisymmetric torque whose magnitude
declined with depth, to mimic forcing from above by differentially
rotating overshooting convection.  He also added a distribution of
random sources of either vertical vorticity, to generate Rossby waves,
or horizontal velocity divergence, to generate gravity waves, the
locations of those sources rotating rigidly (not moving with the
fluid), also with angular velocity $\Omega_0$.  Not surprisingly, wave
drag from dissipation tended to move the differentially forced flow
towards the state of uniform rotation assumed by the externally
imposed grid of sources, which Miesch interpreted as a tendency for
internally generated turbulence to oppose rotational shear.  Evidently
a similar calculation in which the wave sources are contained by the
fluid, the fluid receiving no externally imposed torque other than the
differential forcing from above, would be much more informative.

Propagation and subsequent dissipation of downwardly propagating
gravity waves generated in the convection zone is also a way of
transporting angular momentum through the tachocline, but when such
transport was first mooted 
%RD2 mooted?  Had to look up this sense; I now the other (is moot now)
as a way of modifying the angular velocity
of the Sun 
(\cite{1977ebhs.conf....3G}) 
it was considered not to lead to uniform
rotation.  I have in mind the mechanism now believed to drive the
quasibiennial oscillation of the Earth's atmosphere, a mechanism which
was dramatically demonstrated by \citet{1978JAtS...35.1827P}, and
which arises from the phenomenon that prograde gravity waves,
transporting positive angular momentum, dissipate more readily than
retrograde waves, at a rate which increases with increasing angular
velocity of the fluid, thereby enhancing shear.  However, this
mechanism has not been generally accepted by astrophysicists as being
the dominant determinant of gravity-wave angular-momentum transport
through the Sun.  There are two principal issues: the first is whether
or not the waves are generated to an amplitude great enough for the
transport to be significant, the second is whether they enhance the
shear or suppress it.  These issues have been reviewed by
\citet{2002ESASP.508..577G} and, from a rather different perspective,
by \citet{2007AIPC..948...15C}.  I think it is true to say that they
have not yet been resolved to everyone's satisfaction.  So the debate
will continue.

Finally, I comment on a claim by Forg\'{a}cs-Dajka \& Petrovay (2001, 2002;
\nocite{2001SoPh..203..195F}
\nocite{2002A&A...389..629F}
also \cite{2003SoPh..215...17P}; \cite{2004A&A...413.1143F}) that an
oscillating horizontal magnetic field, presumed to have been generated
by a dynamo in the convection zone, and diffusing into the radiative
interior of the Sun from above, applies a stress on the fluid which
opposes other agents acting on the fluid that tend to transfer
latitudinal differential rotation.  It is argued that it causes the
radiative zone to rotate rigidly, the transition occuring on a
lengthscale 
%RD2 one word?  
%DOG2 It depends how modern you wish to be.  I vote for modernism, 
%DOG2 because I'm still a young man, and retain it as one word.  In any case, 
%DOG2 I don't want length scale with timescale.
%RD3 I love the artist as a young dog.  Use Dopplershift and fluxtube myself
which, with a suitable choice of conditions, could
correspond to the solar tachocline.  Even though the early work was
criticized at the Solar Tachocline meeting held at the Isaac Newton
Institute of Mathematical Sciences, Cambridge in 2004 
(\cite{2007sota.conf....3G}),
for presuming the field presented to the tachocline to be
latitudinally unsheared, and thereby imparting an unsheared Maxwell
stress on the radiative zone which would naturally tend to oppose any
other tendency to shear, the idea in some circles appears to have been
given almost equal credance to the more sophisticated dynamical
discussions (e.g., 
\cite{2007sota.conf...89Z}; 
\cite{2006A&A...457..665B}).
Therefore it behoves 
%RD2 I put a second o in behoves
%DOG2 OED restored for mongrelization avoidance - see e-mail discussion 
%DOG2 on diskussion.
anyone wanting to establish the true dynamical
balance to look at this model more critically.  To this end, Jean-Paul
%RD2 I think this is the only first name in this aper.  > J.-P.?
%DOG2  Thank you for pointing that out.  It is because it occurs anecdotally.
%DOG2  For consistency, I have added Ekman's first name.
Zahn and I have initiated a simple calculation to assess the effect of
the convection-zone shear on the Maxwell stresses, and to estimate the
outcome for a Forg\'{a}cs-Dajka and Petrovay model.  The idea is to
imagine the dynamo field to be generated by a suitable imposed current
distribution in the convection zone, and to calculate the resulting
magnetic field assuming the convection zone to be rotating as is
observed, endowing the fluid in the convection zone with a uniform
scalar turbulent magnetic diffusivity.  We have not yet reached an
agreed conclusion concerning the dynamical response of the
nonturbulent fluid beneath.  But what we have established is that if
the radiative interior were initially rotating uniformly, the stress
transported would have been such as to induce latitudinal differential
rotation in the same sense as that of the convection zone, as one
would expect.  Whether the subsequent evolution would induce a
latitudinal shear that is intense enough to be tested by seismology
remains to be seen.

\begin{acknowledgement}
  I am very grateful to Paula Younger for typing the manuscript, to
  G\"unter Houdek and Jesper Schou for helping with the figures, and
  to Rob Rutten for conversion to the publisher's format.
\end{acknowledgement}

%% References
%%%%%%%%%%%%%%%%%%%%%%%%%%%%%%%%%%%%%%%%%%%%%%%%%%%%%%%%%%%%%%%%%%%%%%%%%%%%
\begin{small}

\bibliographystyle{rr-assp}       %RR hacked from aa.bst
%%\bibliography{gough}  

%RD2 below the inserted .bbl file made by bibtex but edited by me

\end{small}
\end{document}